\documentclass{aa}

\usepackage{graphicx}
\usepackage{natbib}

\bibpunct{(}{)}{;}{a}{}{,}

\usepackage{txfonts}

\begin{document}

   \title{Gamma-ray bursts from tidally spun-up Wolf-Rayet stars?}

   \author{R.G. Detmers\inst{1,2}
          \and
          N.Langer\inst{1}
          \and
	  Ph. Podsiadlowski\inst{3}
	  \and
	  R.G. Izzard\inst{1}
	  }

   \offprints{R.G.Detmers}

   \institute{Astronomical Institute, University of Utrecht, Postbus 80000, 3508 TA Utrecht, The Netherlands \email{r.g.detmers@sron.nl}
         \and
              SRON National Institute for Space Research, Sorbonnelaan 2, 3584 CA Utrecht, The Netherlands 
         \and
              Department of Astrophysics, University of Oxford, Keble Road, Oxford OX1 3RH}

   \date{}

 
  \abstract
   {The collapsar model requires rapidly rotating Wolf-Rayet
   stars as progenitors of long gamma-ray bursts. However, Galactic Wolf-Rayet stars
   rapidly lose angular momentum due to their intense stellar winds.}
   {We investigate whether the tidal interaction of a Wolf-Rayet star with a compact object in a binary system 
   can spin up the Wolf-Rayet star enough to produce a collapsar.}
   {We compute the evolution of close Wolf-Rayet binaries, including tidal 
   angular momentum exchange, differential rotation
   of the Wolf-Rayet star, internal magnetic fields, 
   stellar wind mass loss, and mass transfer. The Wolf-Rayet companion is approximated 
   as a point mass. We then employ a population synthesis code to infer the occurrence
   rates of the various relevant binary evolution channels.}
   {We find that the simple scenario --- i.e., the Wolf-Rayet star being tidally spun up
   and producing a collapsar --- does not occur at solar metallicity
   and may only occur with low probability at low metallicity.
   It is limited by the widening of the binary orbit induced by the strong Wolf-Rayet wind
   or by the radius evolution of the Wolf-Rayet star that most often leads to
   a binary merger. The tidal effects enhance the merger rate of 
   Wolf-Rayet stars with black holes such that it becomes
   comparable to the occurrence rate of long gamma-ray bursts.
}
   {}


   \maketitle
%

\section{Introduction}			\label{intro}

   Although our understanding of long gamma-ray bursts (GRBs) has
   increased significantly since they were first discovered, it is
   still not clear what their progenitors exactly are.  From studies
   of host galaxies it emerged that they occur in or near star-forming
   regions and that several are associated with energetic Type Ic
   supernova \citep{Hjorth03}. These supernovae are thought to stem from
   to the explosion of a massive Wolf-Rayet (WR) star, although the
   signatures of a WR type progenitor have been unambiguously found in the
   afterglow of only one GRB \citep{Marle05}.
   
   The most widely used model for the formation of GRBs is the
   collapsar model \citep{Woosley93a}.  In this model the core of a
   massive, fast rotating star collapses into a black hole. An
   accretion disk is formed around the black hole if the core has
   enough specific angular momentum, i.e. $j$ $\gtrsim
   3\times10^{16}\,$cm$^2\,$s$^{-1}$ \citep{Macfayden99}. The
   remainder of the core is accreted onto the black hole and a highly
   relativistic collimated outflow is produced which releases a large
   amount of energy ($10^{51}\,$ergs).  If the star has no hydrogen
   envelope, the light crossing time is less or comparable to the
   duration of the accretion. In this case a GRB can be formed along
   with a Type Ib/c supernova. Thus, rapidly rotating WR stars
   are required to produce a collapsar.
  
   Earlier work on GRB progenitors \citep{Petrovic05a,Hirschi06} has
   shown that stellar models without a magnetic field can have enough
   specific angular momentum in their cores to produce a GRB within
   the collapsar model.  The single star models in which magnetic
   fields according to \citet{Spruit02} were included had too little
   specific angular momentum in their cores due to the increased
   core-envelope coupling, which leads to a spin-down of the core
   \citep{Petrovic05a, Heger05}.  The reason why magnetic fields are
   considered to be important are the low observed rotation rates of
   young neutron stars and white dwarfs.  In order to reproduce those
   low rates, magnetic torque need to be included \citep{Heger05, Suijs08}.  
   \citet{Petrovic05a} also considered binary models in
   which the secondary star was spun up to close to critical rotation
   due to mass transfer from the primary.  In their model, initially
   the core of the companion was spun up due to magnetic core-envelope
   coupling, but the same mechanism decreased the core angular
   momentum by almost a factor of 100 once the star had reached core
   helium burning.  Therefore, either one has to consider stars with a
   low metallicity which have a lower mass loss rate, or GRBs at
   solar metallicity require a more exotic binary channel.
   The first option has been considered recently by \citet{Yoon05} and
   \citet{Yoon06} and the second option is discussed here.
   
   We investigate here whether tides in a close binary system can spin up a
   WR star enough so that it can form a collapsar.  The magnetic
   fields inside the star enforce close to rigid rotation during core
   helium burning, so that angular momentum added to the envelope due
   to the tidal interaction is expected to be transported to the core
   of the WR star.  In order to have a sufficiently strong tidal
   interaction for spin-up to occur, the orbital period needs to be
   smaller than about 24 h. This restricts potential companions to
   compact stars, helium stars, or low mass main-sequence stars.  While the
   number of double-WR star systems is too small to be significant
   \citep{Vrancken91}, \citet{Heuvel07} showed that for main sequence
   companions, the obtained spin-up is insufficient to produce a GRB.
   Therefore, we focus on compact companions in the following sections.

   Currently, only one compact WR binary is known in our Galaxy, namely Cyg
   X-3. The system has a period of 4.8 h and a period derivative in
   the range of $1-2 \times 10^{-6} $ yr$^{-1}$ \citep{Lommen05}.
   Cyg X-3 has long been known as an X-ray binary.  \citet{Kerkwijk96}
   found that the primary is a WR-star, probably of the WN spectral type.  
   The exact mass of the WR star is unknown, but \citet{Stark03} have placed an upper limit on the mass of 7.3 M$_{\odot}$ 
   using observations taken with \textit{Chandra}. 
   Although the nature of the companion is still not clear, \citet{Kerkwijk96} assume it is a neutron star of 1.4 M$_{\odot}$, while
   \citet{Stark03} place an upper limit of 3.6 M$_{\odot}$ on the compact object mass.
   
   We assume a mass of 10 M$_{\odot}$ for the WR star and 1.4 M$_{\odot}$ for the compact object in our Cyg X-3 case study.
   The mass loss rate is highly uncertain, and estimated by various
   authors to be between $0.5 \times 10^{-6}$ M$_{\odot}$ yr$^{-1}$\ --\ 
   $2.9 \times 10^{-4}$ M$_{\odot}$ yr$^{-1}$ \citep{Lommen05}.
 
   Systems similar to Cyg~X-3, and those considered further on in this
   paper, can only be formed through common-envelope evolution. We
   use Cyg X-3 as a case-study for the tidal spin-up process to
   determine the effect of this process and to see whether it is
   capable of spinning up the WR star enough to meet the collapsar
   criterium.  The remainder of this paper is organised as follows.
   In Sect. 2 we explain our physical assumptions as well as our
   numerical methods.  Our binary evolution models are decribed in
   Sect. 3, while Sect. 4 contains our population synthesis results.
   Finally, our conclusions are discussed in Sect. 5.

\section{Methods and physical assumptions}		\label{methods}
   
   Our stellar models are calculated with a hydrodynamic stellar
   evolution code \citep{Petrovic05a}.  Magnetic fields and the
   transport of angular momentum due to magnetic torque are included
   \citep{Spruit02}, as well as the effects of the centrifugal force
   on the stellar structure, chemical mixing and transport of angular
   momentum due to rotationally induced hydrodynamic instabilities,
   and enhanced mass loss due to close-to-critical rotation.  The
   stellar wind mass loss of the WR star is calculated according to
   \citet[labelled WR0 from now on]{Hamann95} for WR stars with log
   $L/L_{\odot}$ $>$ 4.5:

   \begin{equation}		\label{HamannWR1}
     \log  (\dot{M}{\mathrm{_{WR}} }/ M\mathrm{_{\odot}} $ yr$^{-1}) = \epsilon \times (-11.95 + 1.5 \log {L/L}_{\odot})   , 
   \end{equation}
   and \citet{Hamann82} for helium stars with log  $L/L_{\odot} < 4.5$: 

   \begin{equation}		\label{HamannWR2}
    \log (\dot{M}_{\mathrm{_{WR}}} / M\mathrm{_{\odot}} $ yr$^{-1}) = \epsilon \times (-35.8 + 6.8 \log {L/L}_{\odot})   , 
   \end{equation}
   with $\dot{M}{\mathrm{_{WR}}}$ the mass loss rate of the Wolf-Rayet star and
   $\epsilon$ is a multiplication factor used to modify the mass loss rate.
   
   When a star appoaches critical rotation, the mass loss of the star
   will be enhanced such that over-critical rotation is prevented. To
   achieve this, we follow the prescription of \citet{Langer97}:
   \begin{equation}		\label{mdotcrit}
   \dot{{M}} = \dot{{M}}({V_{\rm rot}} = 0) \left( \frac{1}{1 -  {V_{\rm 
rot}}/ {V_{\rm crit}}}\right)^{0.43}    ,     
   \end{equation}
   where
   
   \begin{equation}		\label{vcrit}
   {V_{\rm crit}} = \sqrt{\frac{GM}{R} (1 - \Gamma)}\   .
   \end{equation}
   Here ${V_{\rm rot}}$ is the rotational velocity of the star,
   $\dot{{M}}$ is the mass loss rate of the star, $\Gamma= \kappa L/(4
   \pi cGM)$ is the Eddington factor, and $\kappa$ is the opacity
   coefficient.  We apply three different rates of mass loss, both
   based on Eqs.~(1) and~(2), using $\epsilon=0.1$, $\epsilon=1/3$,
   and $\epsilon = 1$.  We call these mass loss rates WR1, WR2, and
   WR3, respectively.  The WR2 rate roughly agrees with the mass loss
   rate for Galactic WR stars \citep{Nugis00}.  We chose the
   first rate to study the effect of a lower mass loss rate on the
   evolution of the system.  This essentially captures the situation
   of a lower initial metallicity, since the mass loss rate of a
   WR star is dependant on its metallicity \citep{Vink05},
   while all other properties depend very little on it.
   
   For a given WR mass loss rate, one can define an angular momentum
   loss timescale, i.e. the time it takes for the star to lose most of
   its angular momentum to the stellar wind,
   
   \begin{equation}			\label{tau_ang}
   \mathrm{\tau_{ang}} = \frac{{J_{\rm WR}}}{\dot {J_{\rm WR}}} = \alpha \frac{{M_{\rm WR}}}{\dot{{M_{\rm WR}}}}     ,
   \end{equation}
   where ${M_{\rm WR}}$ is the mass of the Wolf-Rayet star, ${J_{\rm WR}}$ is the total spin angular momentum of the
   star and $\alpha \simeq 0.1$ for efficient internal angular-momentum transport, 
   since angular momentum loss from the wind
   is about 10 times faster than the mass loss \citep{Packet81,Langer98}.
  
  There are two different mechanisms to synchronize a star with the
   orbit, the equilibrium tide and the dynamical tide. Although both
   mechanisms create a tidal bulge which causes the binary to become
   synchronized, they do so on different timescales. The time it
   takes for a star to synchronize with the orbit if it has a
   convective envelope is defined as \citep{Zahn77}:
   
   \begin{equation}		\label{tau_eq}
   \mathrm{\tau_{sync}} \propto  q^{-2} \left(\frac{d}{R}\right)^{6}   .
   \end{equation}  
   where $q = {M_{\rm cc}} / {M_{\rm WR}}$ is the mass ratio of the
   binary, with ${M_{\rm WR}}$ the total mass of the WR star and ${M_{\rm cc}}$ the mass of the compact companion.
   $d$ the orbital separation and $R$ the radius of the star
   to be synchronized.  We choose to use this ansatz for the
   equilibrium tide for determining $\mathrm{\tau_{sync}}$, although
   it was originally designed for convective stars, for two reasons.
   First, in the radiative envelopes of rapidly rotating stars, as
   considered here, rotation is supposed to produce a high level of
   turbulence, implying that turbulent viscosity dominates over the
   radiative viscosity. Second, \citet{Toledano07}, applying a
   kinematic model of tidal interaction to early type stars, found that
   the tidal time scale in these models indeed scales as $(d/R)^6$.
   
   The angular momentum exchange between the orbit and the WR spin is
   computed in the following way \citep{Wellstein}. If $\Delta J$
   is the amount of spin angular momentum that is added to the star
   in the considered time step $\Delta t$ due to tides then,
   
   \begin{equation}		\label{deltalang}
   \Delta J = ({J_{\rm WR}} - {J_{\rm sync}})(1 - {e^{- {\Delta
   t}/{\mathrm{\tau_{sync}}}} }) ,
   \end{equation} 
   where ${J_{\rm sync}}$ is the spin angular momentum the star
   would have when the system is synchronized.
   
   Angular momentum $\Delta J$ is added to the outer layers at
   every time step. Angular momentum transport processes
   (i.e. magnetic torque) redistribute the angular momentum over the
   whole star. Magnetic torque is strong enough to keep the
   star close to rigid rotation during core helium burning. In this
   way the whole of the WR star is spun up due to the tidal
   interaction.
   
   To determine the outcome of the common-envelope evolution we calculate
   the final orbital separation and compare that to the CO-core radius
   and the Roche-lobe radius of the CO-core. We do this using the energy
   equation for common envelope evolution \citep{Webbink84,deKool90}:
  
   \begin{equation}		\label{ce}
     \frac{{a_{\rm f}}}{{a_{\rm i}}} = \frac{{M_{\rm core}}{M_{\rm cc}}}
     {{M_{\rm WR}}} \frac{1}{{M_{\rm cc}} + 2 {M_{\rm env}} {a_{\rm i}} / ({\eta_{\rm CE}} {\lambda R_{\rm L}})},
   \end{equation}
     
  where ${a_{\rm f}}$ is the final orbital separation after the CE
  phase and ${a_{\rm i}}$ the orbital separation at the moment when
  Roche Lobe overflow (RLOF) starts.  ${M_{\rm core}}$ is the mass of the CO-core of the WR
  star and ${M_{\rm env}}$ the mass contained in the envelope of the WR star. 
  The efficiency parameter ${\eta_{\rm CE}}$ is set to 1 in our
  calculations and $\lambda$ is determined by the binding energy of the
  envelope, which is extracted from our models.

\section{Binary evolution models}	\label{models}
		
   We first compute tailored models for Cyg X-3 to determine the role
   of tidal interaction in this system. We assume a
   compact object mass of 1.4 M$_{\odot}$ (i.e. take a neutron star),
   chose a WR-star mass of 10 M$_{\odot}$ and set the initial orbital
   period to 4.8 hrs.  The mass loss rates for our Cyg X-3 models are
   based on the \citet{Hamann95} rate, but with $\epsilon =1$ (model
   WRa), $\epsilon =0.1$ (WRb) and $\epsilon =0.01$ (WRc, see Table
   \ref{Table:1}). Their evolution is described in the next
   subsection.
   
   \begin{table}
\caption{Table 1. Cyg X-3 models.} 
\label{Table:1}      
\centering                          
\begin{tabular}{c c c c c }        
\hline\hline                 
 $\mathrm{M_{WR}}$ [$\mathrm{M_{\odot}}$]& $\mathrm{M_{CC}}$ [$\mathrm{M_{\odot}}$] & $\mathrm{P_{i}}$ [h] & $\epsilon$ & model \\    
\hline                        
   10 & 1.4  & 4.8 & 1    & WRa \\
   10 & 1.4  & 4.8 & 0.1  & WRb \\
   10 & 1.4  & 4.8 & 0.01 & WRc \\      
\hline                                   
\end{tabular}
\end{table}
   
\subsection{Cyg X-3 models: a Case Study}   	\label{cygx3}

   The three Cyg X-3 models which we have calculated show the effects
   of different mass loss rates on the tidal interaction and
   evolution of the system.  The model sequences end either at carbon
   depletion in the core or when the primary fills its Roche-lobe.  In
   all three cases the tidal interaction transfers angular momentum
   from the orbit to the WR star, but the WR star is only spun up if
   the mass loss rate is not too high, i.e. if $\mathrm{\tau_{sync}}$
   $<$ $\mathrm{\tau_{ang}}$.  This is the case for models WRb and
   WRc, but for model WRa the mass loss rate is too high and thus
   $\mathrm{\tau_{ang}}$ too small.

\begin{figure}[tbp]
   \includegraphics[angle=90,width=9cm]{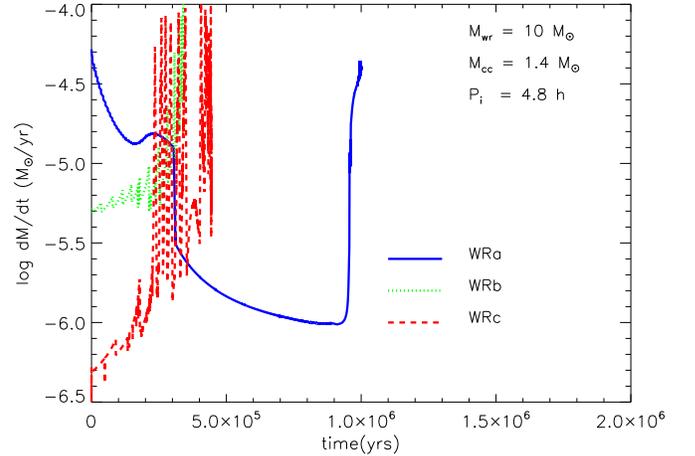}
    \caption{\label{fig:mdot} mass loss rate as a function of time for
       the three Cyg X-3 models: WRa (blue line, solid) with
       $\epsilon$ = 1, WRb (green line, dotted) with $\epsilon$ = 0.1
       and WRc (red line,striped) with $\epsilon$ = 0.01
       (cf. Table~1). }
   \end{figure}

   Figure \ref{fig:mdot} shows the mass loss rates for models WRa, WRb
   and WRc as function of time.  The WRa mass loss rate decreases
   throughout the evolution and only increases sharply at $9.5 \times
   10^{5}$~yr at core helium exhaustion, because the luminosity
   increases greatly when the WR star starts its He-shell burning.
   The WRb and WRc mass loss rates are initially so low that the
   tides spin up the stars to critical rotation, which occurs here
   well before the WR stars fill their Roche lobes since their
   luminosities are near their Eddington luminosities (cf. Eq. \ref{vcrit}). 
   The high luminosity of the WR star has the consequence that the star can approach critical rotation for $\omega_{\mathrm{WR}}$ $<$ $\omega_{\mathrm{orb}}$.
   Once at critical rotation, the mass loss rate is determined by the
   mechanical constraint to avoid over-critical rotation \citep[cf.]{Langer98}. 
   While this is a numerically (and perhaps physically)
   unstable situation \citep[cf. Fig.~3 in]{Langer98}, the corresponding
   large oscillations of the mass loss rate have no consequences for
   the evolution of the system, as long as the long-term average of
   the mass loss rate is still well defined, as is the case here
   (cf. Sect. \ref{wrb}). As the orbital period decreases, the
   time-averaged mass loss rates of systems WRb and WRc increase in
   the course of evolution.

\begin{figure}[tbp]
   \includegraphics[angle=90,width=9cm]{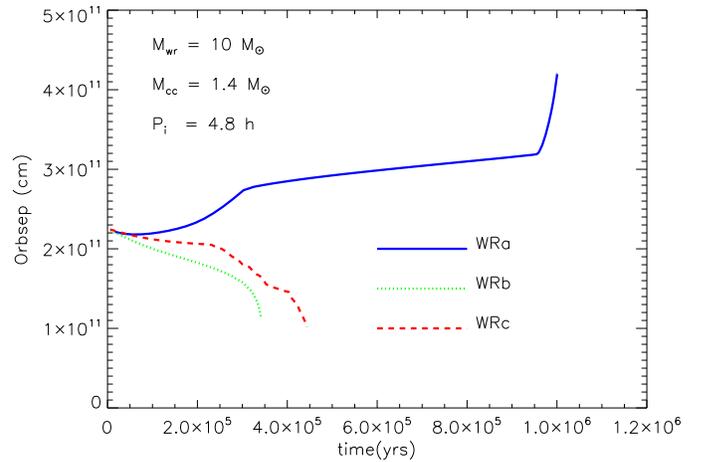}
    \caption{\label{fig:orbits} Orbital separation as a function of
       time for our 3 different Cyg X-3 models. Plotted are model WRa
       (solid line), model WRb (dotted line) and the model WRc (dashed
       line). }
   \end{figure}

   The time evolution of the orbital separation of these three
  sequences is shown in Figure \ref{fig:orbits}. The orbital
  separation of models WRb and WRc decreases, while it increases for
  model WRa.  The evolution of model WRa is dominated by its mass
  loss. The two kinks which occur at $3 \times 10^{5}$ and $9.6 \times
  10^{5}$ yr can be traced back to the changes in the mass loss rate
  at those times (see Figure \ref{fig:mdot}).  In models WRb and WRc
  the initial mass loss rate is low, so that the tidal interaction
  spins the star up and angular momentum is thus transferred from the
  orbit to the WR star. That means that the orbit shrinks and this
  process continues until the WR star starts to fill its Roche-lobe.
  
\subsubsection{Sequence WRa}	\label{wra} 
   Model WRa reaches carbon depletion in the core, although the mass
   of the WR star is only 2.2 M$_{\odot}$ at that time.  The specific
   angular momentum of model WRa does not change significantly during
   the core helium burning phase, implying that the wind induced loss of spin angular momentum is
   compensated by tidally induced angular momentum gain. At the end
   of core carbon burning, the specific angular momentum drops
   to very low values in the core. This is due to the overwhelming
   effect of the increased mass loss, which brings the WR star out of
   tidal locking.

\begin{figure}[tbp]
   \includegraphics[width=9cm]{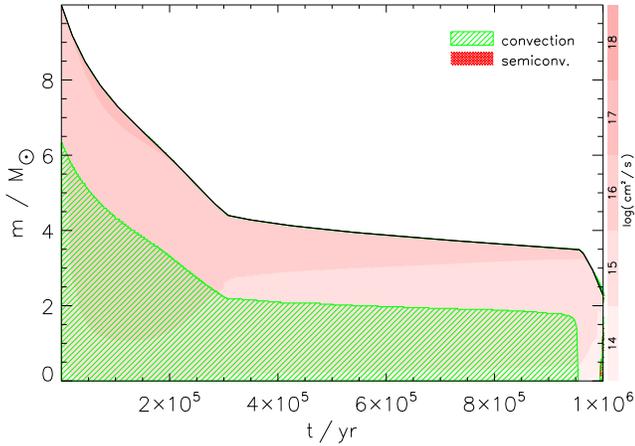}
    \caption{\label{fig:convwra} Kippenhahn diagram of the specific
       angular momentum for the WR star in model WRa. The hatched and
       crossed areas indicate convection and semi-convection
       respectively. The red color indicates the amount of specific
       angular momentum, where a darker red indicates larger specific angular
       momentum. }
   \end{figure}

   The internal specific angular momentum distribution during the evolution of the system can be seen 
   in Figure \ref{fig:convwra}. The angular momentum in the
   core stays roughly constant throughout the evolution, except for an initial increase during the first 40\,000\,yr 
   due to tidal spin-up, and a large decrease after helium shell burning has started at $9.5 \times 10^{5}\,$yr.
   When 50\% helium is left in the core, the star has already lost more than half of its 
   mass due to its strong stellar wind.
   The WR star will not form a black hole because the CO core mass is too low (below 2 M$_{\odot}$), 
   but it will most likely form a neutron star.

\subsubsection{Sequences WRb and WRc}		\label{wrb}
\begin{figure}[tbp]
   \includegraphics[width=9cm]{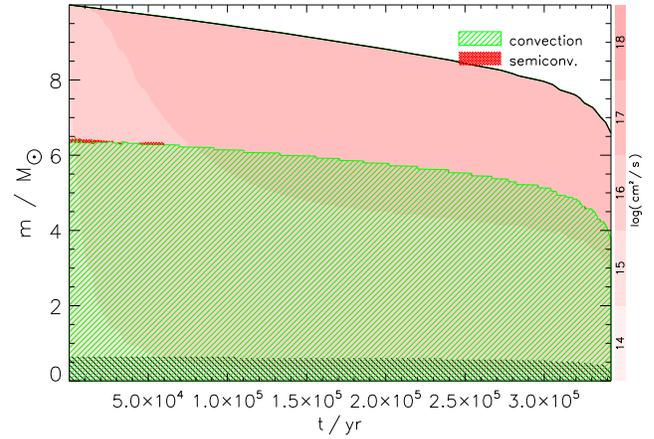}
    \caption{\label{fig:convwrb} As Figure \ref{fig:convwra} except this is for model WRb. The transport of
       angular momentum from the envelope to the core can clearly be
       seen after 80000 yr. }
   \end{figure}

   Model WRb has an entirely different evolution. The star is spun up due to tidal interaction and
as a consequence its specific angular momentum increases.  At the end
of the calculation, the specific angular momentum at 3 M$_{\odot}$ is
10$^{17} \mathrm{cm^{-2}} \mathrm{s^{-1}}$. Figure \ref{fig:convwrb}
shows the internal specific angular momentum distribution during the
evolution. Because the orbital separation decreases with time, the star
remains tidally locked and is spun up throughout the whole
evolution. Figure \ref{fig:convwrb} also shows that despite the
oscillations of the mass loss rate as function of time (Fig.~1), the
time average mass loss rate is well defined, and in fact dictated by
the need to avoid faster than critical rotation at the stellar
surface.  Even though the star loses about 3.5 M$_{\odot}$ the star is
still spinning rapidly at the moment RLOF occurs.  The outcome of the
RLOF phase will most likely be a merger, since the mass transfer
process is unstable because of the large mass ratio.

   The evolution of model WRc is similar to that of model WRb, except
   that the orbital shrinking is less severe in this model. Although
   this may seem strange at first because the mass loss rate is
   lower and thus the tidal interaction should be more effective in
   spinning up the star, the stronger mass loss in model WRb actually
   helps to decrease the orbital separation. This is because the
   system is tidally locked in both models such that any
   angular-momentum loss from the star comes from the orbit
   instead. This effect can be compared to magnetic braking in
   low-mass stars with a convective envelope.  Since the mass loss
   rate is higher in model WRb, the angular momentum loss is also
   larger and the orbit shrinks faster.
   
\begin{figure}[tbp]
   \includegraphics[width=9cm]{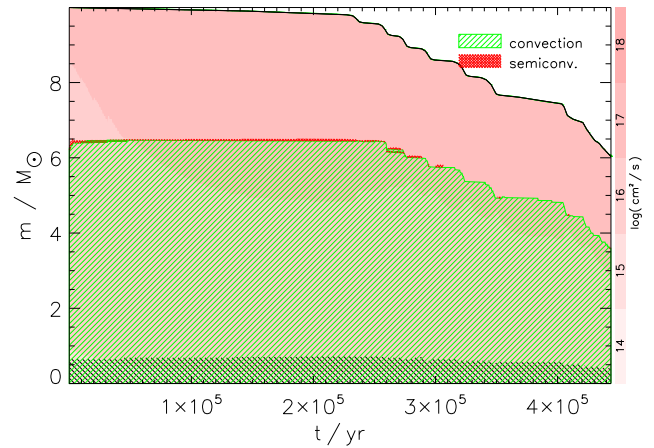}
    \caption{\label{fig:convwrc}
      As Figure \ref{fig:convwra} except this is for model WRc.}
   \end{figure}

 Figure \ref{fig:convwrc} shows the internal specific angular
 momentum as function of time.  Again, the increase of specific
 angular momentum can be seen by the increase of the dark red
 area in the star.  Interestingly, the final mass of the WR star is
 smaller in model WRc than in model WRb. This is due to several
 effects.  First, although the initial mass loss rate is different for
 models WRb and WRc, the effect of the spin-up is that the WR star
 approaches critical rotation. The mass loss rate increases according
 to Eq. \ref{mdotcrit}. The initial value of the mass loss rate does
 not matter once the star approaches critical velocity, which is
 why models WRb and WRc have an almost equal average mass loss rate.
 Secondly, the orbital decay of model WRc is less severe and so the
 point at which the WR star fills its Roche-lobe is reached at a later
 time, so the mass of the WR star is smaller when RLOF starts. The
 likely outcome of this RLOF is again a merger due to unstable mass
 transfer, as is the case for model WRb.

\subsubsection{Summary}
   
In the three Cyg X-3 models discussed above, either the spin-up works
and the orbit shrinks (models WRb and WRc) or the mass loss rate is
too high and the orbit widens (model WRa). The WR star is initially spun up by the tidal
interaction in model WRa, but after 50000 yr the star starts to lose
angular momentum due to the strong stellar wind.  Models WRb and WRc
also spin-up initially, but retain their angular momentum throughout
their evolution. The angular momentum of the orbit always decreases,
due to the mass loss and the tidal interaction. All three systems synchronize their orbits
very quickly and remain synchronized throughout.
      
From these three models it is clear that there is a threshold value
for the mass loss rate which determines the outcome of the system.
Either the mass loss rate is too high so the WR star spins down and the
orbit widens, or the rate is low enough so that the WR star is spun
up. The mass loss rate is determined not by the properties of the WR
star, but rather by the need of sufficient angular-momentum loss to
not exceed the $\Omega$-limit.
Considering the limiting case of the mass loss rate approaching zero, it is clear that the tidal interaction is strong enough to spin
up the star and thus the orbit would shrink. The star would then
eventually approach critical rotation, and the mass loss is enhanced to
a value which is high enough to keep the star from reaching critical
rotation. So the spin-up works as shown in the WRb and WRc systems,
but both systems merge in the end.

\subsection{Parameter study}	\label{parameter}

It is clear that a study of only Cyg X-3 is insufficient to
investigate the whole parameter space. We therefore set up a grid
of binary models with a mass of the WR star between 6 and 18
M$_{\odot}$. Our models all have solar metallicity, and we take a
zero-age helium main sequence star (ZAHeMS) as our WR star.  We use
different masses for the compact object: 1.4 M$_{\odot}$, 3.0
M$_{\odot}$ and 5.0 M$_{\odot}$, i.e. assume either a neutron-star or a
black-hole companion. While larger compact object masses are
possible, systems with such are expected to contribute at most 10 $-$ 20 \% to the total number of helium stars plus compact
object systems. The initial orbital separation is chosen such that the synchronization timescale
$\mathrm{\tau_{sync}}$ is about equal to the stellar wind induced
angular momentum loss timescale $\mathrm{\tau_{ang}}$. The other
chosen initial separations are twice and thrice this equilibrium
value. For the 5 M$_{\odot}$ companion we chose an extra set of
models, with an initial orbital separation of 0.5 times the
equilibrium value.
   
\begin{figure*}
\centering
\includegraphics[angle=0,width=17cm]{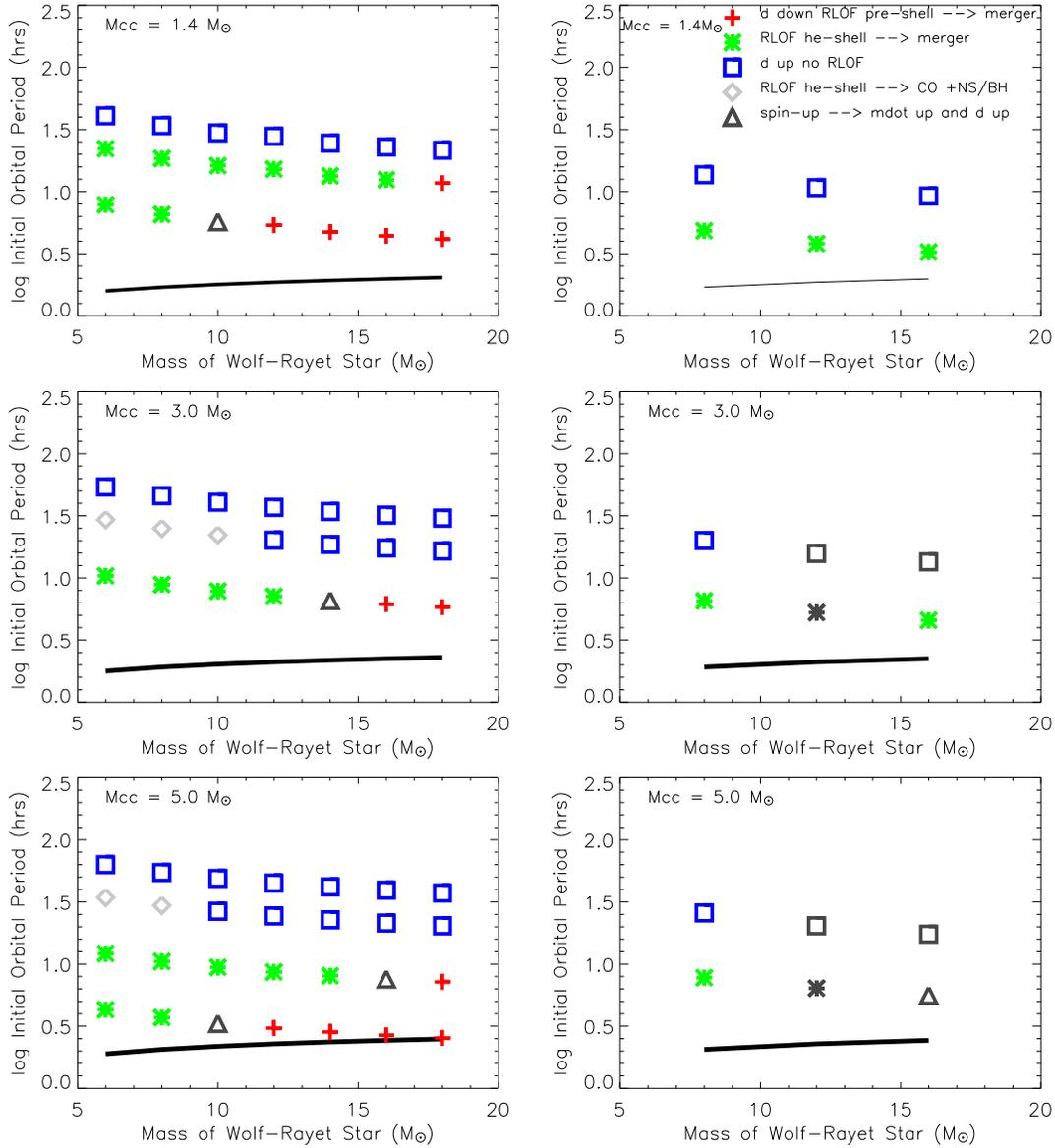}
    \vspace{-2.0cm}
\caption{The outcome of the evolutionary calculations for helium
star-compact object binaries using the WR1 mass loss rate (left
column) and the WR2 mass loss rate (right column). The companion
mass is shown in the upper left corner of each plot. We take a $1.4$
M$_{\odot}$ companion (neutron star), a $3.0$ M$_{\odot}$ companion
(black hole) and a $5.0$ M$_{\odot}$ companion (black hole).
Evolution outcomes are: no RLOF and no spin-up (squares), spiral-in
and RLOF during helium burning (crosses), RLOF during He-shell burning
resulting in a merger (stars), RLOF during He-shell burning resulting
in a CO-core + compact object in a close binary (diamonds), and
spin-up then increased mass loss and orbital widening
(triangles). The solid line indicates where the WR star would
immediately fill its Roche-lobe, which gives the lower boundary for
possible initial periods. }
      \label{fig:grids}
\end{figure*}

As shown for the Cyg X-3 models above, if the spin-up works then the
orbital shrinking will most likely lead to RLOF. To investigate
whether we can have tidal interaction to spin-up the WR star without
ensuring RLOF we expanded our calculations to several model grids.
We use the two different mass loss rates labelled as WR1 and WR2 (see
section \ref{methods}).  Each of the plots in Fig \ref{fig:grids}
shows the mass of the WR star and the initial period for each system.
We computed the evolution of each system and gave each different
outcome a separate symbol in the plot.  We found 5 different outcomes
for the binary system.  

The first and easiest to understand is the case where the initial orbital period is large. 
The tidal interaction is weak in these systems and mass loss dominates the evolution. The
binary widens and the WR star spins down. Depending on the mass of the
WR star it will form either a black hole or a neutron star, but in
neither case is it spinning fast enough to be considered a GRB
progenitor.  

The second and third type of evolution occur when the tidal
interaction is not strong enough to spin-up the WR star, so mass loss
still widens the orbit. However, the system remains compact enough
for the WR star to fill its Roche-lobe during He-shell burning.
Unstable mass transfer starts at that moment and the result is a
common-envelope phase (the second one in the evolution of the binary).
We have found two possible outcomes for the common-envelope phase,
either a merger, or the binary survives the common-envelope phase and
ends up as a CO core with a compact object in a very small orbit.       

The fourth type is the case where the tidal interaction
is strong enough to spin up the WR star. The orbits shrinks and the WR
star fills its Roche-lobe during core He-burning. If the mass ratio
is too far from unity, the mass transfer is unstable and the result is
a merger of both stars. When the compact companion is a neutron star
the result will be a Thorne-\.Zytkow like object \citep{Thorne75}. In the case of a
black hole companion, the WR star will be accreted onto the black hole
and a GRB may occur (we will discuss this option in the next section).
    
In the last type of evolution, the tidal interaction is initially
strong enough to spin up the WR star to close to critical rotation and
the orbit shrinks. Due to the WR star approaching critical rotation,
the mass loss rate increases, which reduces the radius of the WR star, resulting in a weaker tidal interaction. Eventually the system is no longer
tidally locked. The orbit widens again and the star continues to lose
a large amount of mass. The end result is the same as for the systems with a
large initial period.

We do not make a full grid calculation for the WR2 mass loss rate
models, as the main effect of a higher mass loss rate is a shift of
the borderlines between the various types of evolution in the WR1
grids to lower initial periods. The WR2 models are similar to the WR1
models, but the mass loss is higher and thus the initial equilibrium
period is smaller. This is because $\mathrm{\tau_{ang}}$ is smaller,
which means that $\mathrm{\tau_{sync}}$ also has to be smaller in
order to have an equilibrium situation.
    The only potential collapsar progenitors are those systems which
survive the common-envelope phase as a CO star with a compact
companion in a close orbit, if the CO star is spinning fast enough and
is massive enough to form a black hole. The CO core mass of a 10
M$_{\odot}$ WR star is 6.0 M$_{\odot}$, which may be close to the
borderline value for forming a black hole. We have one possible
system in our grid for which this could be the case, namely the
10 M$_{\odot}$ WR star with a 3.0 M$_{\odot}$ companion and an inital
period of $23.15$ h (see Table \ref{Table:2}).
    Out of the 70 systems we calculated with the WR1 mass loss rate, 19
systems may allow GRB production. The most important conclusion drawn
from these models is that there is no model in which only tidal
spin-up works, whilst avoiding RLOF or a merger. So our initial idea
of having only tidal spin-up is unlikely to happen at these mass loss
rates and thus at solar metallicity.
   Tidal interaction is very important for triggering the RLOF or a
merger, thus indirectly contributing to the formation of a possible
GRB. If tidal interaction is weaker than we have assumed here,
then the equilibrium period will shift to a smaller initial orbital
period, the same effect as a higher mass loss rate. If the tidal
interaction is stronger than we have assumed, then the initial orbital
period will shift upwards, the same effect as a lower mass loss
rate.  We did not change the strength of the tidal interaction, but
making the tidal interaction stronger could be one way of making tidal
spin-up work at solar metallicity, although the increase would have to
be significant because of the high mass loss rates at solar
metallicity.
    
We have investigated the possibility of tidal spin-up in a close
WR-compact object binary leading to a collapsar. Our results show
that it is very hard to have tidal spin-up while also avoiding a RLOF
or merger event. At the same time however the tidal interaction is the
cause of the pre He-shell burning mergers. 
These mergers may or may not produce a GRB, which is unknown at the moment.
So indirectly tidal interaction may lead to the formation of a GRB in these cases. 
Even if these sources would not produce a GRB, they may appear as a transient source on the sky.
So our results may have significant observable implications, given the fact that our community is 
currently undertaking big efforts to investigate transient sources at various wavelengths.

At lower metallicity (and lower mass loss rates), the range in periods at
which tidal interaction is strong enough to spin up the WR star is
larger. This is because the `equilibrium period' shifts to a larger
initial period, because $\mathrm{\tau_{ang}}$ is larger and thus also
$\mathrm{\tau_{sync}}$ can be larger. There is a limit to which
lowering the mass loss rate makes a difference, because if the WR star
approaches critical rotation due to the spin-up, the mass loss also
increases and is determined by the angular momentum loss needed to
avoid exceeding the critical rotation rate. So the inital mass loss
rate may only be important in the initial evolution of the system.
One way in which there could be only tidal spin-up would be if the
initial mass loss rate is low and the initial orbital period is large
enough so that the WR star does not approach critical rotation until
the end of its evolution. In that way the WR star would have no time
to spin down again and it will have enough angular momentum in its
core to form a collapsar. So at lower mass loss rates, having only
tidal spin-up is still a fine-tuning process, which means that it is
unlikely that this scenario will be a major GRB production
channel. Lowering the mass loss rate, i.e. assuming lower metallicity,
will have the effect that tidal interaction leading to a RLOF or
merger occurs more frequently, since the range of periods in which
this occurs is wider than at solar metallicity.
        
\section{Population synthesis}

To estimate the birth rate of these kind of systems, we have performed
a population synthesis study using the model of \citet{Hurley02}.
A $100^3$ grid in $M_1$, $M_2$ and $d_\mathrm{init}$ with grid limits
$13<M_1/\mathrm{M}_\odot<55$, $6.5<M_2/\mathrm{M}_\odot<40$ and
$60<d_\mathrm{init}/\mathrm{R}_\odot<1370$, which were chosen after
searching a wide parameter space for progenitor systems, was searched
for systems which match our GRB progenitors. We used the initial mass
function of \citet*{Kroupa93} for the primary, a flat distribution in
$q=M_1/M_2$ between 0 and 1 and a separation distribution flat in
$\log d_\mathrm{init}$ between $3$ and $10^4$ $\mathrm{R}_\odot$, a
Maxwellian supernova kick velocity with dispersion $190 $~$
\mathrm{km/sec}$, solar metallicity ($Z=0.02$), mass loss according to
\citet{Hurley02}, circular orbits, common envelope parameters
$\alpha=1.0$ and $\lambda=0.5$ or $0.05$, compact object (post-SN)
masses according to \citet{Belczynski02} and the star formation rate
given in \citet{Hurley02} of one binary with M$_{1}$ $>$ 0.8 M$_{\odot}$ per year.
    
\begin{table}
\caption{Table 2. Formation rates for each possible GRB progenitor type, for 
$\lambda = 0.5$.}             
\label{Table:4}      
\centering                          
\begin{tabular}{c c c c}        
\hline\hline                 
 scenario & type & fate & birthrate [yr$^{-1}$]\\    
\hline                        
   A & He-shell RLOF      & CO-BH merger & 5.64 $\times$ $10^{-6}$ \\
   B & pre He-shell RLOF  & He-BH merger & 3.83 $\times$ $10^{-5}$ \\
   C & CO + BH            & collapsar?   & 1.39 $\times$ $10^{-7}$ \\      
\hline                                   
\end{tabular}
\end{table}

The formation rates of each of our potential GRB progenitors are given
in Table \ref{Table:4}.  We only consider systems that have a black hole as
a companion.  These systems can be grouped into three categories, each
with a different evolutionary path.  
Group A consists of systems that avoid a RLOF phase during the core helium burning phase of the WR-star
evolution, but experience RLOF and a merger during the consecutive
expansion of the envelope when helium shell burning starts.  
Group B contains the systems which have a strong tidal interaction during core
helium burning so that RLOF begins prior to the helium shell burning
expansion of the star. These systems all end up as mergers.  
We also have one scenario, Group~C, which we consider a possible collapsar
progenitor scenario: The systems that survives the RLOF during helium
shell burning and end up as CO star with a black hole companion. Here,
the CO star is likely spun up by the tides and the remaining life
time may be too short for mass loss or expansion of the WR star to be
significant.
    
The systems in Group~B have a significantly higher formation rate than
systems in the the other two groups. A comparison of the formation
rate of these systems to the average formation rate of GRBs per
galaxy, which is $10^{-5}$ yr$^{-1}$ according to \citet{Postnov00},
shows that these systems could account for a significant fraction of
the GRBs or transient sources.
   
As with all population synthesis studies our results suffer from
uncertainties in the input distributions and physics. Of particular
note for the current work is the highly uncertain common-envelope
evolution and associated free parameter combination $\alpha \lambda$.
We chose those values to match previous studies, such as
\citet{Hurley02}, but we vary $\lambda$ to show its considerable
effect. A higher value for $\lambda$ leads to a less tightly bound
common envelope and a higher chance of envelope ejection --- thus
enhancing the chance to form close WR star plus compact object
binaries.  Reducing $\lambda$ to $0.05$ reduces the rate by a factor
of about 10. This can be understood since a lower $\lambda$ means a
more tightly bound envelope, i.e. less systems are able to eject the
envelope in the CE phase and merge.  The compact object mass
distribution (NS/BH) is also quite uncertain but we have no better
prescription than that used here.
   
\section{Conclusions}

This paper shows that it is not a simple matter to spin up a WR star
through tidal interaction and thereby produce a long gamma-ray burst.
While the spin-up process itself may work, either the evolution of the
binary orbit or of the radius of the WR star prevent the desired
result, in almost all cases. Only at low metallicity, where the WR
star winds may be weak and the orbit can thus be more stable, can this
scenario work for a limited and rather insignificant fraction of
the parameter space.

This negative result does not exclude something interesting 
happening to WR stars with a close companion: most systems with orbital
periods below $\sim 20\,$h lead to a merger. In principle, the
companion could be a compact object, a helium star, or a main sequence
star. As the spin-up scenario fails, only the latter seems interesting
in the context of gamma-ray burst formation \citep[although see][]{Fryer05}, in particular when the compact object is a
black hole. While the merger of a helium star with a black hole or
neutron star does not form a collapsar, it has been proposed that such
events do produce long gamma-ray bursts \citep{Fryer99}. However, detailed models of such mergers which can demonstrate
their ability to produce a GRB are still missing. From the binary
models presented above, we would expect two types of these events,
i.e. a merger during or after core helium buring.  Both may have
rather long time scales compared to the average time scale of long
gamma-ray bursts.  It is also unclear whether these mergers can
produce an explosive event resembling a Type~Ib/c supernova.
Even if these mergers do not produce a GRB, they may be observed as a transient source in the sky.

In summary, the main product of close WR binaries with compact companions
is a helium star-compact object merger --- not a collapsing and rapidly rotating 
WR star. The occurance rate of these events may be compatible with
that of long / soft gamma-ray bursts.  

\begin{acknowledgements}
 We are grateful to Arend-Jan Poelarends for excellent technical assistance, as well as Marten van Kerkwijk for discussion concerning Cyg X-3. 
 RGI thanks NWO for his current fellowship in Utrecht.
\end{acknowledgements}

\bibliographystyle{aa}
\bibliography{bibfiles}

\appendix

\section{Grid models setup}

Table \ref{Table:2} gives an overview of our grid of models using the WR1 mass loss rate, the models using the
WR2 rate are shown in Table \ref{Table:3}.

\begin{table}
\caption{Table 3. Binary model properties for the WR1 mass loss rate.}             
\label{Table:2}      
\centering                          
\begin{tabular}{c c c c c c }        
\hline\hline                 
 $\mathrm{M_{WR}}$ [$\mathrm{M_{\odot}}$]& $\mathrm{M_{CC}}$ [$\mathrm{M_{\odot}}$] & $P_{1} $[h] & $P_{2} $[h] & $P_{3} $[h] & $P_{4}$ [h] \\    
\hline                        
   6 & 1.4  & 7.83 & 22.16 & 40.70 & - \\      
   8 & 1.4  & 6.53 & 18.47 & 33.94 & - \\
   10 & 1.4 & 5,71 & 16.16 & 29.68 & - \\
   12 & 1.4 & 5.14 & 14.55 & 26.73 & - \\
   14 & 1.4 & 4.72 & 13.36 & 24.55 & - \\ 
   16 & 1.4 & 4.40 & 12.45 & 22.86 & - \\      
   18 & 1.4 & 4.14 & 11.72 & 21.52 & - \\
   \hline
   6 & 3.0  & 10.40 & 29.41 & 53.99 & - \\      
   8 & 3.0  & 8.84 & 25.00 & 45.93 & - \\
   10 & 3.0 & 7.83 & 23.15 & 40.69 & - \\
   12 & 3.0 & 7.12 & 20.13 & 36.98 & - \\
   14 & 3.0 & 6.58 & 18.61 & 34.20 & - \\ 
   16 & 3.0 & 6.16 & 17.44 & 32.03 & - \\      
   18 & 3.0 & 5.82 & 16.48 & 30.28 & - \\
   \hline
   6 & 5.0  & 12.14 & 34.34 & 63.08 & 4.29 \\      
   8 & 5.0  & 10.49 & 29.69 & 54.55 & 3.71 \\
   10 & 5.0 & 9.41 & 26.62 & 48.91 & 3.33 \\
   12 & 5.0 & 8.63 & 24.41 & 44.85 & 3.05 \\
   14 & 5.0 & 8.04 & 22.74 & 41.77 & 2.84 \\ 
   16 & 5.0 & 7.57 & 21.41 & 39.34 & 2.68 \\      
   18 & 5.0 & 7.19 & 20.33 & 37.35 & 2.54 \\
   \hline                                   
\end{tabular}
\end{table}

\begin{table}
\caption{Table 4. Same as Tab \ref{Table:2}, except the WR2 mass loss rate is used. }             
\label{Table:3}      
\centering                          
\begin{tabular}{c c c c c }        
\hline\hline                 
 $\mathrm{M_{WR}}$ [$\mathrm{M_{\odot}}$]& $\mathrm{M_{CC}}$ [$\mathrm{M_{\odot}}$] & $P_{1} $[h] & $P_{2} $[h] & $P_{3} $[h] \\    
\hline                        
   8  & 1.4  & 4.83 & 13.67 & - \\
   12 & 1.4  & 3.81 & 10.77 & - \\
   16 & 1.4  & 3.26 & 9.21  & - \\      
   \hline
   8 & 3.0   & 6.54 & 20.02 & - \\
   12 & 3.0  & 5.27 & 15.76 & - \\
   16 & 3.0  & 4.56 & 13.48 & - \\      
   \hline
   8 & 5.0   & 7.77 & 25.84 & 2.75 \\
   12 & 5.0  & 6.39 & 20.35 & 2.26 \\
   16 & 5.0  & 5.60 & 17.41 & 1.98 \\      
   
   \hline                                   
\end{tabular}
\end{table}

\section{Angular momentum evolution}

\begin{figure}[htbp]
   \includegraphics[angle=90,width=9cm]{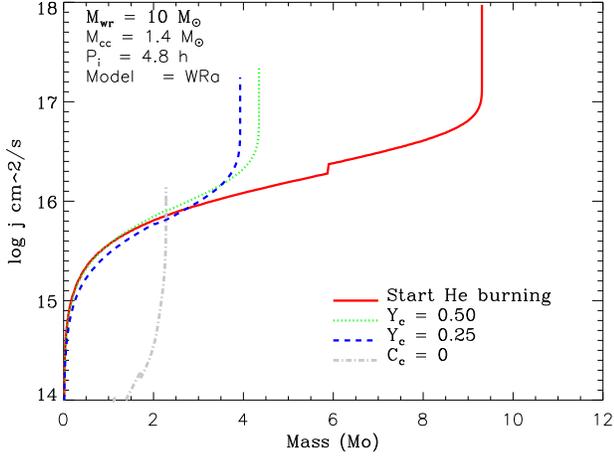}
    \caption{\label{fig:jwra}
       Specific angular momentum as a function of mass coordinate for our Cyg X-3 model WRa (1.4 M$_{\odot}$ companion). Profiles are for the start of core helium burning (solid line), 50\% He-depletion in
       the core (dotted line), 25\% He-depletion in the core (dashed line) and the last model calculated when carbon is depleted in the core (dot-dashed line).}
   \end{figure}

\begin{figure}[htbp]
   \includegraphics[angle=90,width=9cm]{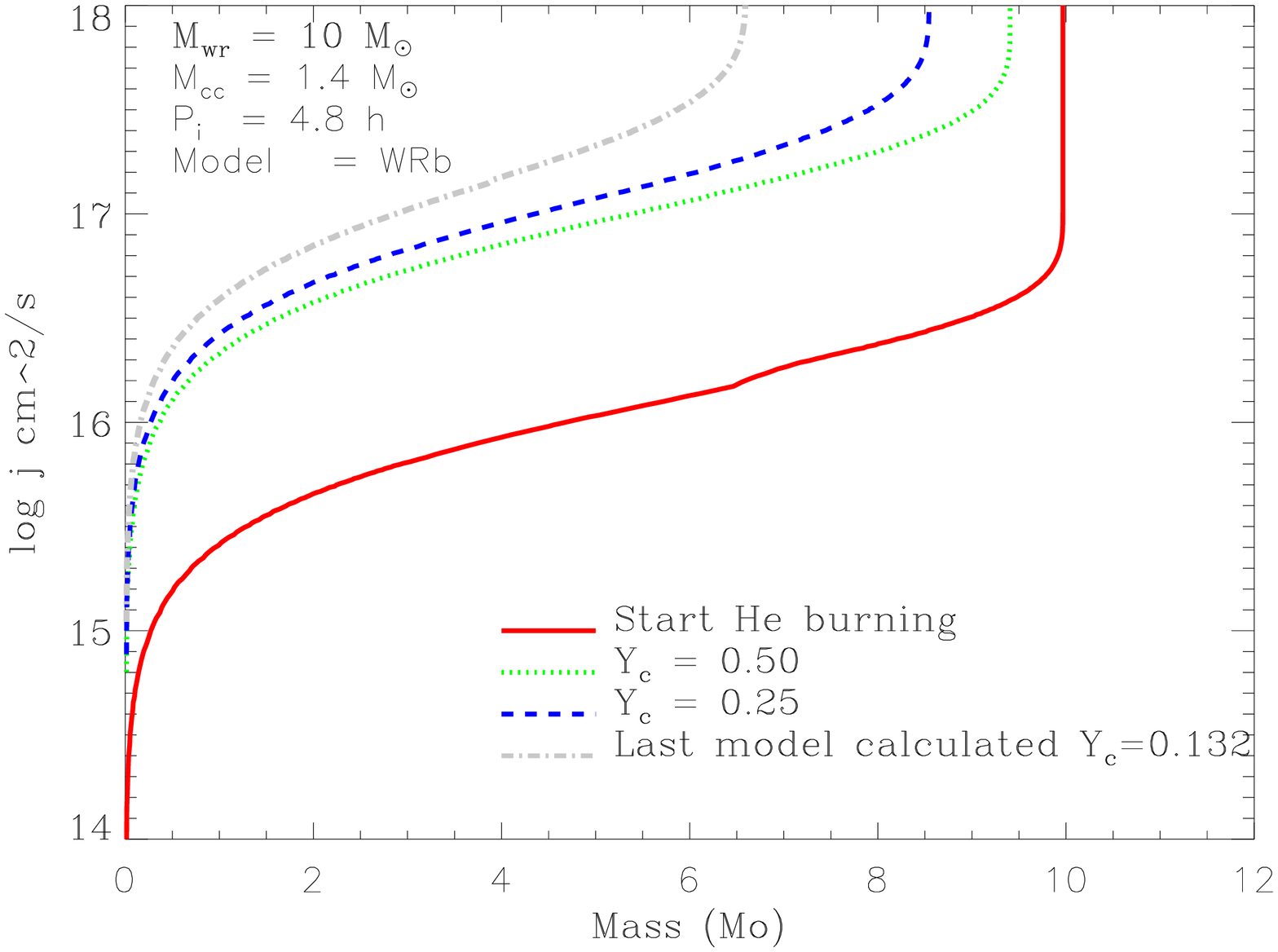}
    \caption{\label{fig:jwrb} Specific angular momentum as a function
       of mass coordinate for Cyg X-3 model WRb (1.4 M$_{\odot}$
       companion). Profiles are for the start of core helium burning
       (solid line), 50\% He-depletion in the core (dotted line), 25\%
       He-depletion in the core(dashed line) and the last model
       calculated when RLOF started (dot-dashed line).}
   \end{figure}

\begin{figure}[htbp]
   \includegraphics[angle=90,width=9cm]{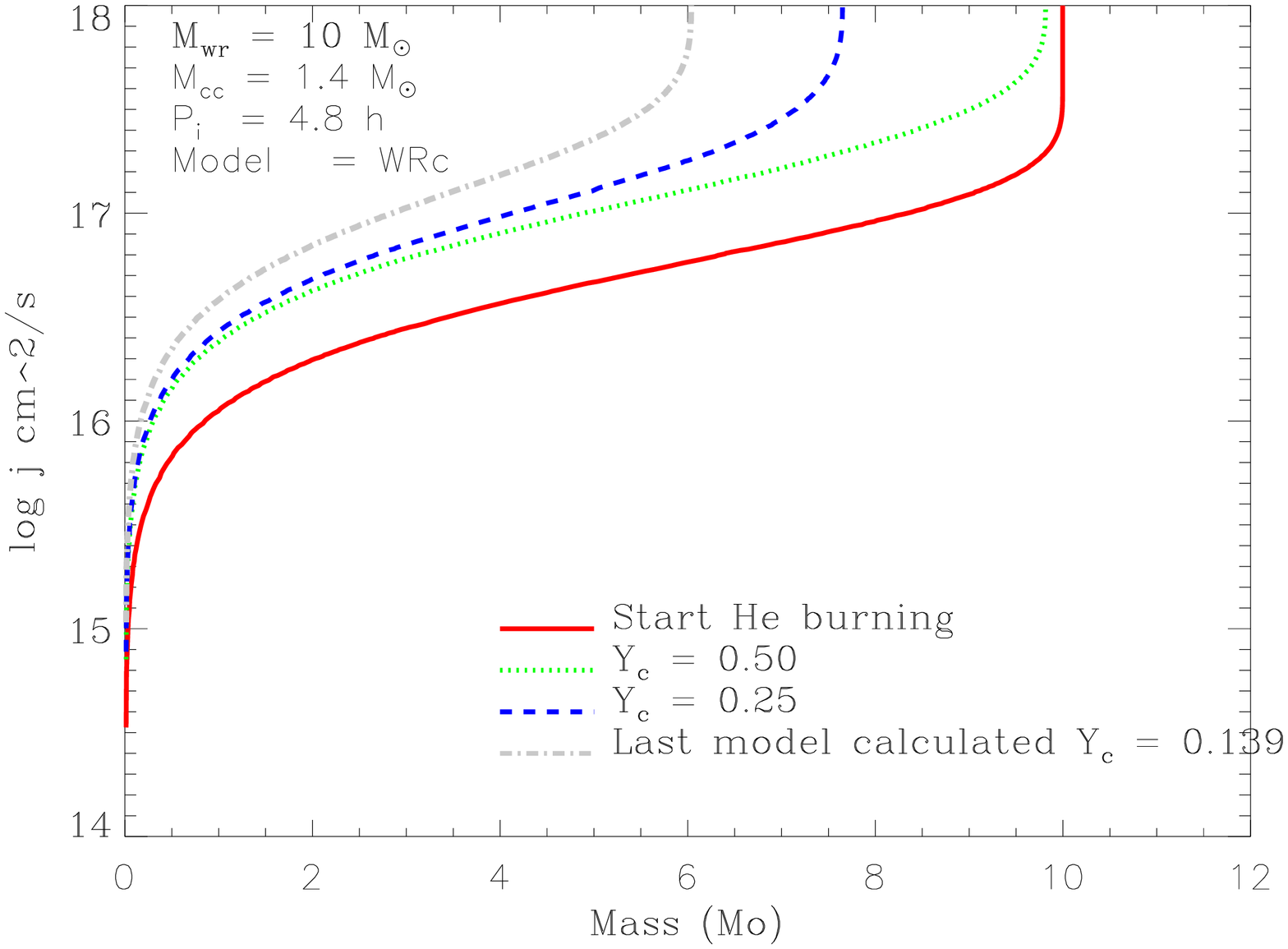}
    \caption{\label{fig:jwrc}
       Specific angular momentum as a function of the mass coordinate for Cyg X-3 model WRc (1.4 M$_{\odot}$ companion). Profiles are given for the start of core helium burning (solid line), 50\% He-depletion in
       the core (dotted line), 25\% He-depletion in the core (dashed line) and the last model calculated when RLOF started (dot-dashed line).}
   \end{figure}

The detailed specific angular momentum profiles of the WR stars for each model (WRa, WRb and WRc) can be seen in Figures \ref{fig:jwra}, \ref{fig:jwrb} and \ref{fig:jwrc}.
These all show that during core helium burning the specific angular momentum profile inside the star does not change significantly.
Also clear is that the specific angular momentum of model WRa, does not increase during the evolution of the system, while for models WRb and WRc the spin-up clearly works, i.e. the specific angular momentum increases.

\begin{figure}[htbp]
   \includegraphics[angle=90,width=9cm]{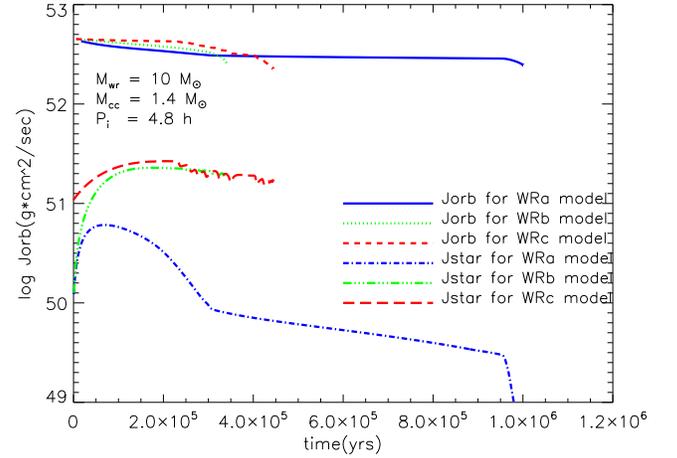}
    \caption{\label{fig:jorbits}
   Orbital and stellar angular momentum as a function of time for our three Cyg X-3 models. The orbital angular momentum (solid lines) is plotted for model WRa (blue, solid), model WRb (green,dashed)
   and model WRc (red,dotted). The stellar angular momentum (dotted lines) has the same coloring for the different models.}
   \end{figure}

\begin{figure}[htbp]
   \includegraphics[angle=90,width=9cm]{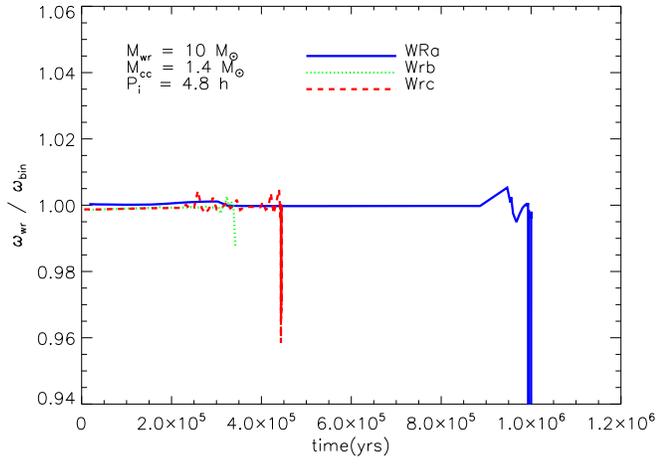}
    \caption{\label{fig:angorb}
   The ratio of angular velocity and orbital angular velocity $\mathrm{\omega_{wr}}$ / $\mathrm{\omega_{bin}}$ for model WRa (solid blue line), model WRb (dotted green line) and 
   model WRc (dashed red line).}
   \end{figure}

Figures \ref{fig:jorbits} and \ref{fig:angorb} show the evolution of the orbital angular momentum and the degree of synchronization ( $\omega_{\mathrm{WR}}$ / $\omega_{\mathrm{orb}}$ ).
The angular momentum of the orbit always decreases, due to either the strong mass loss of the WR star (WRa) or tidal spin-up (WRb + WRc).
The system remains synchronized throughout its evolution till RLOF starts.

\end{document}